\documentstyle[preprint,tighten,eqsecnum,aps]{revtex}

   \newcommand{\be}{\begin{equation}}
   \newcommand{\ee}{\end{equation}}
   \newcommand{\bea}{\begin{eqnarray}}
   \newcommand{\eea}{\end{eqnarray}}
   
   \newcommand{\ket}[1]{\mbox{$|{#1}\rangle$}}
   \newcommand{\bra}[1]{\mbox{$\langle{#1}|$}}
   \newcommand{\braket}[2]{\mbox{$\langle{#1}|{#2}\rangle$}}
   
   \newcommand{\bm}[1]{\mbox{\boldmath$#1$}}
\renewcommand{\d}{\displaystyle}
\renewcommand{\S}{\mbox{$\cal S$}}
\newcommand{\Kr}[1]{\left( #1\right)}
\newcommand{\Kg}[1]{\left\{ #1\right\}}

\newfont{\bld}{cmssbx10}

\begin{document}
\draft

\title{\bf Effective charge-spin models for quantum dots}
\author{John H. Jefferson$^1$ and Wolfgang H\"ausler$^2$\\[5mm]
$^1$DRA, Electronics Sector, St. Andrews Road, Malvern,
Worcs.~WR14 3PS.~U.K.\\
$^2$I.~Institut f\"ur Theoretische Physik, Jungiusstr.~9,
20355 Hamburg, F.~R.~G.}

\date{\today}
\maketitle

\begin{abstract}
It is shown that at low densities, quantum dots with few electrons may be
mapped onto effective charge-spin models for the low-energy eigenstates.
This is justified by defining a lattice model based on a many-electron
pocket-state basis in which electrons are localised near their classical
ground-state positions. The equivalence to a single-band Hubbard model is 
then
established leading to a charge-spin ($t-J-V$) model which for most
geometries reduces to a spin (Heisenberg) model. The method is refined to
include processes which involve cyclic rotations of a ``ring" of neighboring
electrons. This is achieved by introducing intermediate lattice points and 
the
importance of ring processes relative to pair-exchange processes is
investigated using high-order degenerate perturbation theory
and the WKB approximation. The energy spectra are computed from
the effective models for specific cases and compared
with exact results and other approximation methods.
\end{abstract}


\narrowtext

\section{Introduction}
Technological advances in microfabrication with corresponding
reduction in feature sizes has led to renewed interest in the
transport properties of semiconductor submicron structures
both from the fundamental physics point of view and for
possible future applications. In small metallic islands the
Coulomb blockade gives rise to the single-electron effects,
which may be modelled classically by a small intra-dot
capacitance \cite{likharev}. Islands fabricated on the basis of
semiconductors, called {\it quantum dots}, show in addition to
the charging effects discrete energy levels related to
size quantisation \cite{sikorski,heitmann,meirav}. Individual quantum dots
may be fabricated from hetero-structures, which limits the electron
motion to two dimensions, by imposing lateral confinement with
a metallic gate electrode deposited using fine-line (e-beam)
lithography \cite{meirav}. The discrete energy spectra may be measured
in transport experiments at finite voltages \cite{korotkov,weinmann94}
or frequencies \cite{bruder}. The influence of Coulomb
correlations on the excitation energies is difficult to see
optically by far infrared absorption \cite{merktphysica} due to
the generalized version of Kohn's theorem \cite{chaplik90}. The
observation of the quadrupole transitions by using grating couplers
has been suggested \cite{wagner95} as a possible optical means.

Quantum dots have been referred to as `artificial
atoms' \cite{kastner} since the number of conducting
electrons can be smaller than ten or twenty.
At the low electron densities which are experimentally accessible, the
calculation of excitation energies becomes a challenging theoretical
problem. For real atoms, an independent-electron picture correctly
describes the main physics and the Hartree Fock approximation yields
reasonably accurate eigenstates which may be refined in a controlled
way by perturbation theory. This is not generally the case for
semiconducting quantum dots for which even an optimal Hartree Fock
approximation is significantly in error \cite{daniela93a} and can even
give qualitatively incorrect results, such as the wrong spin-multiplet
structure. The reason for this qualitatively different behaviour is
that the electrons in a semiconducting quantum dot are highly correlated,
due to the low effective density and the restriction of the electron
motion to only two dimensions. Many body effects must be taken into
account. This has been done by numerically exact diagonalizations for
systems with a very few electrons ($N\le 4$) \cite{maksym}. In order to
obtain reasonably accurate spectra for systems with
more electrons, approximations must be made. 
For electrons in high
magnetic fields, the low-energy states of maximal spin have been
determined to good accuracy by making a harmonic expansion about
classical minimum energy configurations for circular dots with parabolic
confining potentials \cite{maksym2}. The resulting effective Hamiltonian
is then diagonalised explicitly and
electron antisymmetry is subsequently imposed to determine the allowed
eigenstates. High accuracy is achieved by including a sufficient number
of Landau levels in the basis set.
A complementary technique which has achieved some
measure of success at low electron densities, and is not restricted to
high magnetic fields, is based on many-electron ``pocket'' states 
\cite{haus93}.
Low-energy spectra of systems with up to six electrons have been determined
accurately \cite{pocket} and the agreement with the exact numerical
solutions ($N\le 4$) is good. Unfortunately the computational
effort is likely to become prohibitive for systems with more than ten
electrons and we are again faced with the problem of devising a
reliable approximation method for such cases.

One possible route is suggested by the pocket-state analysis itself
which exploits the permutation symmetry of the wavefunctions. We
notice that in many cases the eigensolutions may be represented
by an effective spin model of a simple form (Heisenberg model).
This apparent equivalence is reminiscent of magnetic insulators
for which the correlated electron problem is known to reduce to a
spin-Hamiltonian, a mapping which may be justified by transformation
theory \cite{chao} or degenerate perturbation
theory \cite{jjlett} as well as, in some cases, by the theory of
permutation groups \cite{rutherford}. The magnetic insulator
problem starts with a lattice model in which the electrons are
localised on atomic-like orbitals and for which electron
correlations are essential, the generic model being the so-called
Hubbard model \cite{hubbard}. The underlying crystalline lattice
establishes sites on which suitable one-electron states are
centred. The question arises as to what circumstances, if any,
might a Hubbard-type model be applicable to the problem of
interacting electrons in a quantum dot, for which the Fermi wave
lengths are much larger than the interatomic spacings and the
underlying crystalline lattice loses its significance. At
extremely low electron densities the long range interaction
energy, according to Wigner \cite{wigner}, creates a crystalline like
ground state which might define new electronic lattice sites.
This crystallization is expected to take place only at electron
densities which are a factor of 20 smaller than the densities
used in experiments. Nevertheless, it has been shown that the
assumption of localised electrons allows the calculation of discrete
low energy excitations as quantum corrections to the Wigner
crystal energies \cite{pocket}. The validity of a Hubbard model
description is not only an interesting question of principle but
has the practical potential of enabling systems of many more
electrons to be dealt with than has hitherto been the case, i.e.\
tens of electrons or more. This would enable the full machinery
of techniques for solving the Hubbard and Heisenberg models to be
applied, for which there has been much progress in recent years
following the discovery of high-temperature superconductivity
and the associated theoretical activity on correlated electron
systems \cite{dagotto}.

In this paper we will show that such a mapping may indeed be
justified for the a priori {\em continuous} problem of
interacting electrons in a quantum dot at low densities. This is
done through the use of pocket-states which are briefly reviewed
in the next section after introducing the basic interacting
electron model. In section~\ref{hubm} we show how the pocket-states
may be used to define a one-electron orthonormal basis with orbitals
localised on a ``lattice", related to the electron
configuration(s) in the classical ground state (determined by
electrostatics). This is then used to construct a tight-binding
Hamiltonian which reduces to the Hubbard model at low electron
densities. Further mapping to an effective (Heisenberg) spin
Hamiltonian or charge-spin ($t-J-V$) model is then performed.

In section~\ref{ring} we point out an essential difference
between quantum dots and the corresponding correlated electron
problem for interacting electrons on a true lattice of atoms,
namely the increased importance of the so-called ``ring" terms in
the former case, which involve cyclic permutations of more than
two electrons. The simple Hubbard model on a lattice
underestimates the magnitude of these ring terms for the dot and
it is shown how the method may be refined by introducing
intermediate lattice points which are unoccupied in the ground
manifold but give rise to the required ring processes through
virtual excitations. In section~\ref{compare} we solve the effective
Hamiltonians for specific examples with $N\le 6$ and compare with
exact numerical and pocket-state results where appropriate.
Finally, in section~\ref{outlook}, we give a summary and discuss the
outlook for this approach in dealing with systems of more electrons and
of obtaining further corrections where necessary.

\section{The Model and Pocket-State Basis}\label{model}
We consider the $\:N$--electron quantum dot described by the Hamiltonian
\bea\label{genmodel}
H&=&\d\sum_{i=1}^N\Kr{\frac{\bm{p}_i^2}{2m}+v(\bm{x}_i)}+
W(\bm{x}_1\ldots\bm{x}_N)\\[3ex]
&&\d W(\bm{x}_1\ldots\bm{x}_N)=
\frac{1}{2}\sum_{i,j\atop i\ne j}w(|\bm{x}_i-\bm{x}_j|)\quad,
\eea
where $\:\bm{x}_i\:$ and $\:\bm{p}_i\:$ are position and momentum
of the $\:i\:$--th electron
in $\:d$--dimensions ($\:d=2\:$ for most quantum dots)
with (effective) mass $\:m\:$ and spin $\:s=1/2\:$.
Neither the one-particle confinement potential $\:v(\bm{x})\:$
nor the interaction $\:w(x)\:$ depend explicitly on spin.

The cases of square well single particle potentials in one
dimension (1D) and in two dimensions (for square geometry) have
been discussed in \cite{pocket}. To be specific, and for comparison
later, we shall also mainly confine ourselves to the 2D square
potential well of square geometry, though the extension to other
geometries is straightforward and, for the present consideration,
the detailed form of $\:v(\bm{x})\:$ is not qualitatively
important. At large mean inter--particle distances, $\:r_{\rm
s}\:$, it becomes energetically favourable for the electron
system to localize its charge density distribution in regions
close to the classical ground-state electron configuration(s)
\cite{wigner,jauregui}, which may be determined by minimizing the
electrostatic energy.

This fact is the motivation for the many-electron pocket-state
basis in which the Hilbert space is restricted to $\:1\le
p\le\nu\cdot N!\:$ basis states $\:\ket{p}\:$ with spatial
representations being defined in configuration space of dimensionality
$\:d\:N\:$. The number of classical minimum energy configurations
$\nu$ may be greater than one in certain symmetric geometries.
(For example, for $N=3$ on a square we have $\nu=4$ since, for
the classical ground-state energy, any one of the four corners
may be unoccupied.) The low-energy levels form a multiplet
which contains $\:\nu\cdot 2^N\:$ states in total (including
Zeeman levels). The excited `vibrational' states scale like a
power law $\:\sim r_{\rm s}^{-\gamma}\:$ with $\:r_{\rm s}\:$
($\:\gamma\:$ being close to $\:3/2\:$) and can be neglected
if the density is not too high. This is due to the scaling
$\:\sim\exp(-\sqrt{r_{\rm s}/r_{\rm c}})\:$ of the low energy
excitations we are interested in here,
where $r_{\rm c}$ is a length scale which characterizes the
transition from the almost non-interacting situation
$(r_{\rm s}\ll r_{\rm c})$ into the qualitatively different regime of
strong correlations $(r_{\rm s}\gg r_{\rm c})$ \cite{haus93}.

In the pocket-state method, the first step is to calculate
energies by assuming equivalent but distinguishable particles and
ignoring their spin. Spin and statistics are subsequently
reestablished by means of group theory.
Tunneling integrals $\:\bra{p}H\ket{p'}\:$ between the localized
many particle (pocket--) basis states determine the low--energy
excitations. The tunneling processes correspond to (correlated)
transitions between different particle arrangements. In the
low--density regime one tunneling integral, $\:J/2\:$, is
exponentially larger than all others. In many cases the dominant
$\:J\:$ corresponds to the exchange of only two particles, which
may be adjacent in real space. Other processes could be
the simultaneous (ring--) exchange of three or more particles,
corresponding to a cyclic permutation.

An important feature of the spectra given by the pocket-state
approximation (and by the effective charge-spin models to be
derived and discussed in the following sections) is that the
fine-structure of the energy spectra depends only on $J$, the
magnitude of which may be estimated semiclassically within the
multidimensional WKB approximation. The ratios between the energy
differences are insensitive to the detailed form of the
inter--electron potential and to $\:r_{\rm s}\:$. An examination
of the spectra shows that these situations may be mapped onto an
effective spin model with again one parameter for each process
(pair exchange, ring) considered. This equivalence will be
justified in the next section. Here we merely make the
observation that for the cases considered by the pocket-state
method, the energy spectrum is the same as that which would be
given by a spin-Hamiltonian. An exception to this are situations
for which more than one classical minimum energy configuration
for the electrons exist ($\:\nu>1\:$), such as are 2 or 3
electrons on a square. We will show in the next section that these
latter situations may be mapped onto a so-called $t-J-V$ model in
which the dominant tunneling process is now related to $t$, the
amplitude corresponding to the jump of a single particle into an empty
`site'. $J$ is again an exchange process involving at least two
particles and $V$ represents the Coulomb repulsion between
electrons on neighboring `sites'.

In general, the pocket-states are correlated electron states
which need not be assumed as a direct product of one--electron
states. However, we may mimic the eigenstates at sufficiently
large $r_{\rm s}$ with a one--electron product basis within the
Heitler London (HL) approximation.
The latter are localised in regions close to the
positions the electrons would have in the classical ground state.
In the next section we show how this one-electron basis may be
used to define an orthonormal basis from which we construct an
effective Hubbard model. This effective Hamiltonian yields
low-energy spectra which are at least as accurate as what would be
given by the corresponding HL approximation using a product basis
of non-orthogonal, one-electron wavefunctions.
We then show in section~\ref{ring} how the method may be extended to
cases where ring processes are important.

\section{Mapping to Hubbard and Charge-Spin Models}\label{hubm}
Let us assume that the pocket basis states \ket{p} are
approximated ``optimally" by non-orthogonal one-electron
wavefunctions $\:\phi_i\:$ centred at position $i$. These
states have the form,
\be\label{product}
\braket{\bm{x}_1,\ldots,\bm{x}_N}{p}\equiv
\phi_{p_1}(\bm{x}_1)\phi_{p_2}(\bm{x}_2)\ldots\phi_{p_N}(\bm{x}_N)
\ee
where $\:\{p_1,\ldots,p_N\}\:$ is the permutation $\:p\:$ of
the sequence $\:\{1,\ldots,N\}\:$. In this section we
orthogonalise these one-electron wavefunctions and construct
antisymmetrised $N$-electron states from which we define the
effective Hubbard model, which is subsequently transformed into a
spin-Hamiltonian. To illustrate the main features of the method
with minimal mathematical complexity, we first consider in detail
the simplest non-trivial case of two electrons in a 1D square
well. The analysis is then generalised to $N$-electrons in one or
two dimensions.

As is well known, for two electrons the antisymmetrised states may be
written as the product of orbital and spin parts, as with the HL
states of the hydrogen molecule. The symmetric (singlet) and
antisymmetric (triplet) orbital states are thus,
\be
\Psi_S=\frac{\phi_1(1)\phi_2(2)+\phi_2(1)\phi_1(2)}
{\sqrt{2(1+s^2)}}
 \label{S}
 \ee
 and
 \be
 \Psi_A=\frac{\phi_1(1)\phi_2(2)-\phi_2(1)\phi_1(2)}
{\sqrt{2(1-s^2)}}
 \label{A}
 \ee
where $s=\bra{\phi_1}\phi_2\rangle$ is the overlap and the arguments
$\:(i=1,2)\:$ abbreviate the coordinate $\:\bm{x}_i\:$ of the
$\:i\:$--th particle. These HL states yield approximate singlet and
triplet ground-state energies: $E_{\rm singlet}=\bra{\Psi_S}H\ket{\Psi_S}$
and $E_{\rm triplet}=\bra{\Psi_A}H\ket{\Psi_A}$, where $H$ is the
Hamiltonian for two electrons given by equation (\ref{genmodel}).

We now transform to orthonormal one-electron states $\psi$ where
\bea\label{trans}
\d\psi_1=\left(\frac{1}{2\sqrt{1+s}}+\frac{1}{2\sqrt{1-s}}\right)\phi_1
 + \left(\frac{1}{2\sqrt{1+s}}-\frac{1}{2\sqrt{1-s}}\right)\phi_2\\[3ex]
\d\psi_2=\left(\frac{1}{2\sqrt{1+s}}+\frac{1}{2\sqrt{1-s}}\right)\phi_2
 + \left(\frac{1}{2\sqrt{1+s}}-\frac{1}{2\sqrt{1-s}}\right)\phi_1\quad.
 \eea
Inverting this transformation and substituting into equations (\ref{S}) and
(\ref{A}) we get,
\be\label{OS}
\Psi_S=\frac{1}{\sqrt{1+s^2}}\Psi_S^{(1)}
+\frac{s}{\sqrt{1+s^2}}\Psi_S^{(2)}
\ee
\bea\label{OSI}
\d\Psi_S^{(1)}=\frac{\psi_1(1)\psi_2(2)+\psi_2(1)\psi_1(2)}{\sqrt{2}}\\[3ex]
\d\Psi_S^{(2)}=\frac{\psi_1(1)\psi_1(2)+\psi_2(1)\psi_2(2)}{\sqrt{2}}
\eea
and
\be
\Psi_A=\frac{\psi_1(1)\psi_2(2)-\psi_2(1)\psi_1(2)}{\sqrt{2}}\quad.
\label{OA}
\ee

We note that $\Psi_A$ (triplet) has the same form as for the
non-orthogonal states given in equation (\ref{A}), i.e.\ it
corresponds to one electron in each orbital. On the other hand
$\Psi_S$ (singlet) consists of two components, one with an
electron on each orbital ($\Psi_S^{(1)}$) and the other with both
electrons in the same orbital ($\Psi_S^{(2)}$). (The latter state
is not allowed for the triplet by the Pauli principle.) This
``double occupation" is thus a direct consequence of
orthogonalisation.

Using these two-electron states composed of orthogonal one-electron
wavefunctions will, of course, yield exactly the same
approximate singlet and triplet energy expectation values.
However, we can obtain a more accurate estimate for the ground-state
energy of the singlet by diagonalising the Hamiltonian in
the Hilbert space defined by the two-electron base states
$\Psi_S^{(1)}$ and $\Psi_S^{(2)}$ (Eqn. \ref{OSI}). This follows from
the variational principle which ensures that diagonalisation of
the Hamiltonian matrix will yield the optimum superposition of
the base states, whereas the HL state (\ref{OS}) is not optimum
in general.

This principle also applies to the general $N$-electron case as
stated in the introduction, i.e.\ diagonalisation of the
Hamiltonian in the restricted Hilbert space defined by base
states (Slater determinants) constructed from orthogonalised
one-electron wavefunctions will yield a more accurate low-energy
spectrum than the simple products of non-orthogonal one-electron
states (\ref{S}) and (\ref{A}).

For the two-electron case the ($2\times 2$) Hamiltonian singlet matrix is,
\be
H_{\rm s}=\left( {\matrix{{2\varepsilon +V+j}&{2\tilde{t}}\cr
{2\tilde{t}^*}&{2\varepsilon+U+\tilde{\jmath}}\cr
}} \right)
\label{singlet}
\ee
where
\bea\label{parameters}
\varepsilon&=&\d\bra{\psi_1}[\frac{\bm{p}^2}{2m}+v]\ket{\psi_1}
= \bra{\psi_2}[\frac{\bm{p}^2}{2m}+v]\ket{\psi_2},\\[3ex]
V&=&\d\int{|\psi_1(\bm{x})|^2 |\psi_2(\bm{y})|^2
 w(|\bm{x}-\bm{y}|)\:{\rm d}^3x\:{\rm d}^3y},\\[3ex]
U&=&\d\int{|\psi_1(\bm{x})|^2 |\psi_1(\bm{y})|^2 w(|\bm{x}-\bm{y}|)
 \:{\rm d}^3x\:{\rm d}^3y}=\int{|\psi_2(\bm{x})|^2 |\psi_2(\bm{y})|^2
 w(|\bm{x}-\bm{y}|)\:{\rm d}^3x\:{\rm d}^3y},\\[3ex]
t&=&\d\int{\psi_1^*(\bm{x})\left[\frac{\bm{p}^2}{2m}+
v(\bm{x})\right]\psi_2(\bm{x})\:{\rm d}^3x},\\[3ex]
\tau&=&\d\int{\psi_1^*(\bm{x})\left[\int{|\psi_1(\bm{y})|^2
w(|\bm{x}-\bm{y}|)d^3y}\right]\psi_2(\bm{x})\:{\rm d}^3x},\\[3ex]
\tilde{t}&=&t+\tau,\\[3ex]
j&=&\d\int{\psi_1^*(\bm{x}) \psi_2^*(\bm{y})w(|\bm{x}-\bm{y}|)
\psi_2(\bm{x}) \psi_1(\bm{y})\:{\rm d}^3x\:{\rm d}^3y},\\[3ex]
\xi&=&\d\int{\psi_2^*(\bm{x}) \psi_2^*(\bm{y})w(|\bm{x}-\bm{y}|)
\psi_1(\bm{x}) \psi_1(\bm{y})\:{\rm d}^3x\:{\rm d}^3y},\\[3ex]
\tilde{\jmath}&=&\mbox{Re}(\xi)\quad.
\eea
Note, $\:j=\xi=\tilde{\jmath}\:$ if the $\:\psi_i\:$ are real.

The triplet energy is,
 \be
 E_t=2\varepsilon+V\quad.
 \label{triplet}
 \ee

Within this restricted manifold of states, the singlet-triplet
Hamiltonian matrix, equations (\ref{singlet}) and (\ref{triplet}),
are equivalent to an effective Hamiltonian,
\bea
H_{\rm eff}&=&\varepsilon \left[ {n_1+n_2}\right]+U\left[
n_{1\uparrow}n_{1\downarrow}+n_{2\uparrow}n_{2\downarrow}\right]+
Vn_1n_2+j\sum\limits_\sigma {c_{1\sigma}^\dagger}c_{2\bar\sigma}^\dagger
c_{2\sigma }^{}c_{1\bar \sigma }^{}\nonumber\\[3ex]
&&\mbox{}+\left[ \sum\limits_\sigma \left(t\:c_{1\sigma}^\dagger
c_{2\sigma}^{}+\tau (c_{1\sigma}^\dagger c_{2\sigma}^{}n_{1\bar\sigma}
+c_{2\sigma}^\dagger c_{1\sigma}^{}n_{2\bar\sigma})
+\frac{1}{2}\xi\:c_{1\sigma}^\dagger c_{1\bar\sigma }^{\dagger}
c_{2\bar\sigma}^{} c_{2\sigma}^{}\right)+{\rm hc} \right]
\label{two}
\eea
where $c^\dagger_{i\sigma}$ is a Fermi creation operator at `site' $i=1,2$
satisfying $\bra{\bm x}c^\dagger_{i\sigma}\ket{\rm vac}=\psi_{i\sigma}({\bm
x})$; $n_{i\sigma}=c^\dagger_{i\sigma}c^{}_{i\sigma}$ and
$n_i=n_{i\uparrow}+n_{i\downarrow}$.

This effective Hamiltonian is very similar to that considered by
Hubbard \cite{hubbard} for a periodic array of one-electron atoms in
the study of the metal-insulator (Mott) transition. Indeed, for
this two-electron case, equation (\ref{two}) is an effective
Hamiltonian for the hydrogen molecule within a restricted Hilbert
space of (effective) 1s-orbitals. We stress, however, that in the
quantum dot case the localised orbitals are fundamentally
different from those in the atomic case in that their very
existence depends on the electron-electron repulsion which is
responsible for localising the electrons near specific points in
real space at low density. For this reason we are not justified,
{\em a priori}, in dropping all but the largest Coulomb term
($U$), as is usually done in the Hubbard model. In particular,
the nearest-neighbor effective Coulomb interaction, $V$, becomes
most important when $\:r_{\rm s}\:$ becomes very large and
equation (\ref{two}) simply reduces to the classical expression
for the ground-state energy, independent of spin, as it should.
This may be seen more clearly by transforming (\ref{two}) into an
effective spin-Hamiltonian which is demonstrated most directly for
this two-electron case by diagonalising the singlet-matrix,
(\ref{singlet}), explicitly to yield eigenenergies:
\be
E_\pm=\frac{1}{2}\left[4\varepsilon +V +U+j+\tilde{\jmath}\pm
\sqrt{(U-V+\tilde{\jmath}-j)^2 + 16|\tilde{t}|^2}\right]\quad.
\ee

Since for the cases of interest (large separation between the electrons
and real one-particle wave functions), $\:U\gg V,j,\tilde{\jmath}\:$ and
$\:j=\tilde{\jmath}\:$, then,
\be
E_+\approx 2\varepsilon+U+j+\frac{4|\tilde{t}|^2}{U-V}
\ee
and
\be
E_-\approx 2\varepsilon +V+j - \frac{4|\tilde{t}|^2}{U-V}\quad.
\ee
Combining these equations with (\ref{triplet}) for the triplet we see that 
the
energy spectrum consists of a singlet-triplet pair at low energies separated 
by
a singlet at energy $\sim U$ higher. Hence the low-energy singlet-triplet
is equivalent to a spin system with effective spin-Hamiltonian,
\be
H_{\rm spin}=2\varepsilon +V +J({\bm s}_1\cdot{\bm s}_2 -1/4)
\ee
where
\be
J=\frac{4|\tilde{t}|^2}{U-V} - j \approx
\frac{4|\tilde{t}|^2}{U} -j
\label{J}
\ee
The first term in equation (\ref{J}), favouring a singlet ground-state,
is sometimes referred to as `superexchange' and is usually larger than
the second `direct' exchange term, $j$.
This superexchange contribution has its origins in the assumed
non-orthogonality of the initial Heitler-London basis $\phi_i$
in (\ref{product}). From the pocket states it is known that they form a
non-orthogonal basis set for the low energy eigenstates \cite{huller80}.

We now consider the general case of $N$ electrons in a quantum dot. Starting
with a suitable product-basis of pocket-states, equation (\ref{product}), we
may, in general, define an orthonormal basis by L\"owdin's method
\cite{lowdin}. (For high-symmetry situations it may be more convenient and expedient to use 
some
other method. For example, in the case of four electrons on a square we have
$\pi/4$ rotational symmetry and we may generate orthonormal states by
forming molecular orbitals, normalising and then transforming back to
localised states. Similarly, for translationally invariant systems we may
transform to running (Bloch) waves, normalise and then transform back to
localised (Wannier) states.) Within the corresponding many-electron Hilbert
space we may express the Hamiltonian in second quantised form.
We are led directly to an effective 'single-band' model of the
type considered by Hubbard, generalising equation (\ref{two}),
\bea
H_{\rm eff}=&&\sum\limits_i \left[ \varepsilon _in_i+
U_in_{i\uparrow}n_{i\downarrow } \right]+\sum\limits_{ij}
V_{ij}n_in_j+\sum\limits_{ij\sigma } j_{ij}c_{i\sigma }^\dagger c_{j\bar
\sigma }^\dagger c_{j\sigma }^{}c_{i\bar \sigma }^{}\nonumber\\
 &&+ \left[ \sum\limits_{ij\sigma }\left(
 t_{ij}c_{i\sigma }^\dagger c_{j\sigma}^{}
 +\tau_{ij} (c_{i\sigma }^\dagger
c_{j\sigma}^{}n_{i\bar \sigma } +c_{j\sigma}^\dagger c_{i\sigma }^{}n_{j\bar
\sigma }) +\frac{1}{2}\xi_{ij} c_{i\sigma}^\dagger c_{i\bar\sigma }^{\dagger}
c_{j\bar\sigma}^{} c_{j\sigma }^{}\right)+{\rm hc} \right]
\label{heff}
\eea
where the indices $\:i\:$ and $\:j\:$ should be distinct. The parameters
$\varepsilon, U, V, j, t,\tau$ and $\xi$ have the same meaning as in
the two-electron case (\ref{parameters}), though it should be noted that
they will, in general, depend on the position of the electron, or pair
of electrons and on separation. In particular the parameters $j, t,\tau$
and $\xi$ related to electron transfers decay rapidly as the distance
between the `sites' $\:i\:$ and $\:j\:$ increases. This allows us
to retain only nearest neighbor terms in the double summations.
As with the case of two electrons, we may eliminate the high-energy states of
$H_{\rm eff}$ to yield an effective spin-Hamiltonian. This may be done using
degenerate perturbation theory \cite{lindgren} or a canonical transformation
\cite{schriefferwolf}, as with the case of a lattice.

Specifically in the situations of two (or higher) dimensional
quantum dots the cases of more than one minimum of the classical
electrostatic energy may occur ($\:\nu>1\:$). Then the quantum
mechanical charge density distribution shows more than $\:N\:$
peaks at large $\:r_{\rm s}\:$, say $\:\tilde{N}>N\:$. The number
of single electron states must be equal to $\:\tilde{N}\:$ in these
cases. $\:\tilde{N}-N\:$ lattice sites will be unoccupied at an instant.
Consider, for example, the case of a square confining potential
in two dimensions. For $N=2$, $\:\tilde{N}=4\:$ and the electrons
will be found mainly near opposite vertices along the diagonal of
the square, due to the Coulomb repulsion. States with two
electrons near adjacent vertices will be of order $V$ higher
in energy (nearest-neighbor Coulomb repulsion energy) and states
with two electrons located near the same vertex will be of order
$U$ higher in energy. Similarly for $N=3$ there will be low-lying
states with electrons located near three of the $\:\tilde{N}=4\:$
vertices with states having two electrons near the same vertex
again being $\sim U$ higher in energy. For $N = 4$ all vertices
are occupied in the ground manifold, with configurations
for which there are two electrons near one vertex being at least
$\sim U$ higher in energy.

In each of these cases the high-energy states, corresponding
to one or more vertices being occupied by
two electrons, may be eliminated from the Hilbert space by
degenerate perturbation theory, where they appear as intermediate
states. More precisely, we first write $H_{\rm eff}$ in the form
$H_{\rm eff}=H_0+H_1$ where $H_0=\sum\limits_i \left[
\varepsilon_in_i+ U_in_{i\uparrow}n_{i\downarrow} \right]$. $H_0$
is diagonal in the basis of product states with the highest-energy
states having two electrons in the same state, being $\sim
U$ higher than the low-energy states \cite{footnote}.
Eliminating these states to
second-order results in an effective `$t-J-V$' Hamiltonian
\cite{tJVref},
\be H_{\rm eff}^{tJV}=
P\sum_{ij}\left[\sum_\sigma (t_{ij}c_{i\sigma }^\dagger c_{j\sigma}^{}
+{\rm hc})+ J_{ij}(\bm{s}_i\cdot\bm{s}_j-1/4)n_in_j+V_{ij}n_in_j\right]P
\label{tJV}
\ee
In this equation $t$ and $V$ have the same meaning as before and
\be
J_{ij}=2|t_{ij}+\tau_{ij}|^2\left[\frac{1}{U_i}+\frac{1}{U_j}\right]-j_{ij}
\quad. \label{J1}
\ee
$P$ is a projection operator which eliminates all base states for
which there are two electrons in the same localised state. (These states
appear as intermediate states and give rise to the `superexchange' in 
equation
(\ref{J1}).) We also note that the $\varepsilon, \tau$ and $\xi$ terms in
equation (\ref{heff}) have disappeared in equation (\ref{tJV}). The former 
has
been dropped since it always gives rise to a constant and the latter
two are precluded by $P$ (as is the $U$-term) since they would always give
rise to double occupation at a `site'.

Equations (\ref{heff}) and (\ref{tJV}) are the fundamental effective
Hamiltonians for interacting electrons at low-density in a quantum dot
and capture the main physics of the problem for most situations. (They
do not, however, include processes in which groups of electrons can
rotate simultaneously, which can be important in some circumstances.
The generalisation to include these so-called ring terms is given in
the next section.) Although these effective models were justified by
starting with the Heitler-London approximation to pocket states using
non-orthogonal one-electron states, we emphasize that they have a
greater range of validity than the initial approximation suggests.
The above methodology enabled us to define a basis set of
product states (Slater determinants) of some one-electron states and
this led directly to effective Hamiltonians operating within this
restricted (incomplete) basis set. However, it is well known
\cite{lindgren}, that higher-lying base states may be accounted for
by perturbation theory. This technique has recently been applied very
successfully to lattice models in which an effective single-band
Hubbard model was derived from a multi-band model using
quasi-degenerate perturbation theory \cite{multi}. We shall not
attempt such a reduction here for the quantum dot problem, but merely
point out that its main effect would be to renormalise the effective
parameters ($\varepsilon_i, U_i, t_{ij}$ etc) and extend the validity
of the resulting models.

Let us apply (\ref{tJV}) to the case of electrons in a square well
in two dimensions, discussed above. For $N=2$ and $N=3$ the $t$-term
will cause electrons to hop from occupied to unoccupied sites whereas
the $J$ and $V$ terms will be effective only when two adjacent sites
are occupied. The presence of the $V$-term ensures that for $N=2$ the
electrons will have lower energy when they are located on diagonally
opposite vertices of the well, as mentioned earlier. For $N=4$ the
$t$-term may be omitted since all four vertices are occupied in the
ground manifold and it would thus give rise to double occupancy, precluded
by $P$. The $V$-term may also be dropped since it gives rise to a constant
contribution. Hence, the sites $\:i\:$ and $\:j\:$ can be restricted to 
nearest
neighbors so that (\ref{tJV}) reduces to a Heisenberg model,
\be
H_J=\sum_{<ij>}J_{ij}\bm{s}_i\cdot\bm{s}_j
\label{hj}
\ee
where $P$ has been omitted for brevity.

Similar considerations apply for $N>4$ and the effective
Hamiltonian will either have the form (\ref{hj}) or (\ref{tJV}),
depending on whether $\:\nu=1\:$ (e.g.\ $N=4,5$) or $\:\nu>1\:$
(e.g.\ $N=2,3,6$). For the latter cases a further reduction will
be possible when some of the states of the $t-J-V$ model are
clearly higher in energy due to the $V$-interaction.

Consider, for example, the case two electrons in a square-shaped
dot for which there are six ways of distributing the two
(indistinguishable) electrons in the four corners, resulting in a
Hilbert space of dimension $4\times 6=24$ when spin is taken into
account. Four of these configurations (16 with spin) are of order
$V$ higher in energy than the remaining two, as shown in
figure~\ref{squararr}. These high-energy base states may be
eliminated from the problem by a further application of
degenerate perturbation theory, resulting in an effective
Hamiltonian which only operates in the Hilbert space of the 8
lowest-energy base states. By a straightforward calculation this
takes the form (apart from an unimportant overall constant):
\be
H_{\rm eff}= \Delta (n_1n_3-n_2n_4)(R_{\pi/2}-R_{-\pi/2})
+J\left[({\bf s}_1\cdot {\bf s}_3-1/4)n_1n_3
+({\bf s}_2\cdot {\bf s}_4-1/4)n_2n_4\right]
\label{delta}
\ee
where $\Delta=2t^2/V$ and $R_\theta$ is a rotation operator,
i.e.
\be\label{rotop}
R_{\pi/2}\ket{\sigma,*,\sigma',*}=\ket{*,\sigma,*,\sigma'}\quad,
\ee
etc. Note that the prefactor ($n_1n_3-n_2n_4$) takes the value
$\pm 1$ depending on whether either `sites' 1 and 3, or 2 and 4 are
occupied. This effective Hamiltonian is easily diagonalised. The
spectrum consists of two singlets and two triplets. We see
immediately that (\ref{delta}) gives zero when operating on a
spin-polarised state and hence the triplets are at zero energy
with the singlets at energies $-J\pm 2\Delta$, in agreement with
what is obtained from the pocket state approximation for this
problem when the exchange of the two electrons is included there
\cite{pocket}. There are thus two
contributions to the `binding energy' of the singlet ground
state, a resonance energy ($-2\Delta)$ and a superexchange
energy ($-J$). It is easy to see that the former will be dominant
since it occurs in second-order and does not involve
double-occupation of a `site'. On the other hand, there are two
contributions to $J$, the second-order term $\sim 4t_{13}^2/U$
and fourth-order terms $\sim t_{12}^4/V^2U$. We may extend
these considerations to a regular polygon with $N$ vertices and
$N/2$ electrons. The effective Hamiltonian will again consist of
rotation operators (of $\pm 2\pi/N$) and exchange terms. The
exchange terms are independent of $N$, whereas the rotation terms
becomes relatively less important since they first occur in
$Nth$-order. In the next section we shall encounter
circumstances where simultaneous rotations of more than two
electrons may be important.

We conclude this section by emphasizing the computational
advantage in this reduction to a spin or charge-spin Hamiltonian.
Thus, for example, the Heisenberg model lies in a Hilbert space
of dimension $2^N$ for $N$ electrons, which may be further reduced by
exploiting other symmetries, e.g.\ conservation of total $S_z$ which
reduces the largest subspace to $N!/((\frac{N}{2})!)^2$ ($S_z=0$ for
$N$ even). This should be compared with
\[
\max_S\Big[\frac{(2S+1)N!}{(N/2+S+1)!(N/2-S)!}\Big]^2
\]
for the largest subspace of the Hamiltonian in the original
pocket-state basis.

The mapping also gives interesting insight into the
nature of the energy spectrum. For those cases which
reduce to a Heisenberg model, the spectrum consists
of spin--multiplets with energy--separation of order $J$ (`bandwidth'
$\sim 2NJ$). On the other hand, for cases which reduce to a
charge-spin model, there are multiplets separated by energies of
order $t$ ($\gg J$ in the same geometry and electron densities), each
of which have a fine structure with energy splittings of order $J$.

The forgoing analysis for electrons may also be extended to a fictitious
system of spin-1/2 charged bosons. For repulsive Coulomb interactions
they also show Wigner crystallization and the pocket-state method
can be applied. This problem was considered by H\"ausler \cite{pocket} in
order to proof the state of highest spin within the multiplet structure of
electrons (fermions) in a quantum dot being the spin polarized state for
$\nu=1$ by exploiting an isomorphism between fermions and bosons using
permutation group theory. We reconsider this problem here to enable
another direct comparison with the pocket-state description.

The case of spin-1/2 bosons is similar to the fermion case in that at
low densities the largest `Coulomb' matrix elements correspond to two
bosons on the same `site'. Neglecting all other matrix elements thus
gives the `boson' Hubbard model:
\be
H=\sum\limits_{<ij>\sigma }(t_{ij}b_{i\sigma}^\dagger b_{j\sigma}^{}+H.c.)+
\frac{U}{2}\sum\limits_{i,\sigma\sigma'}
b_{i\sigma}^\dagger b_{i\sigma'}^\dagger b_{i\sigma'}^{}
b_{i\sigma}^{}
\label{hb}
\ee
Using the usual commutation rules for the Bose-operators yields
\bea
b^\dagger\ket{n}=\sqrt{n+1}\ket{n+1}\nonumber\\
b\ket{n}=\sqrt{n}\ket{n-1}\quad.
\label{crules}
\eea
It follows that:
\[
h_0\ket{\sigma,\sigma'}=U\ket{\sigma,\sigma'}
\]
where
\[
h_0=\frac{U}{2}\sum\limits_{\sigma\sigma'}
b_{\sigma}^\dagger b_{\sigma'}^\dagger b_{\sigma'}^{}
b_{\sigma}^{}
\]

This is the same result as the case of fermions except that with bosons
the two particles may have the same spin.

As with the fermion case, we can now reduce (\ref{hb}) to a Heisenberg or
$t-J-V$ model (depending on whether or not the number of bosons equals the
number of lattice sites). Let us, for simplicity, do this for just two
sites. The extension to the general case may be done by perturbation
theory in the same way as with fermions.

In the `atomic limit' ($t=0$) the eigenstates of (\ref{hb}) with two bosons
consist of either one boson on each site with either spin (energy 0) or
one site unoccupied and the other doubly occupied, again with either spin
(energy = $U$). Using these base states it is straightforward to show
that the singlet state:
\[
\ket{\psi_S}=\frac{\ket{\uparrow,\downarrow}-\ket{\downarrow,\uparrow}}
{\sqrt{2}}
\]
is unaffected by the hopping, i.e.
\[
H\ket{\psi_S}=0
\]
On the other hand the triplet states mix. For example, using
(\ref{crules}), we get:
\[
H\ket{\uparrow,\uparrow}=t\left[ \ket{\cdot,\uparrow\uparrow}
 +\ket{\uparrow\uparrow,\cdot}\right]\quad,\]
\[
H\ket{\uparrow\uparrow,\cdot} = U\ket{\uparrow\uparrow,\cdot}
 +\sqrt{2}t\ket{\uparrow,\uparrow}
\]
etc. This leads directly to the triplet Hamiltonian matrix:

$$\left[ {\matrix{0&{2t}\cr
{2t}&U\cr
}} \right]$$
with lowest eigenvalue:
\[
E=\frac{U-\sqrt{U^2+16t^2}}{2}
\]
This is exactly the same result as for the fermions except that the
positions of the (lowest) singlet and triplet levels are reversed.
It follows that {\em the lowest singlet and triplet is equivalent to a
Heisenberg model with ferromagnetic exchange of the same magnitude as the
case of fermions.} As mentioned above, this may be generalised to
any number of sites by degenerate perturbation theory
resulting in a Heisenberg model in which $J \rightarrow -J$
when compared with the fermion case, leading to an `inverted'
spectrum in agreement with the permutation group analysis
described in the second reference of \cite{pocket}.

In a similar fashion, cases for which the number of bosons is less
than the number of sites (such as 2 or 3 on a square) may be reduced
to a $t-J-V$ model. However, whilst the sign of $J$ is reversed
relative to the case of fermions, the sign of $t$ remains the same
and the resulting spectrum is no longer a simple inversion.

\section{Ring terms}\label{ring}
In the section~\ref{hubm} it was shown that the $t-J-V$ model may
be reduced to an effective Hamiltonian which only involves the
lowest-energy base states. This effective Hamiltonian then
contains operators which rotate some or all of the electrons
simultaneously (cf.\ (\ref{delta})).
The possibility of such a collective motion of a subset or all of
the electrons is much more general and will in fact occur for any
geometry. These processes are negligible within the framework of
the usual lattice models but this is not always the case in
quantum dots or other semicondutor based nano-structures
where
such `ring' terms can become dominant. As we shall see they may
be included in a lattice model by introducing intermediate
lattice points which are unoccupied in the ground manifold.

Consider, for example, the case of an equilateral triangular dot
containing three electrons, as shown in figure~\ref{triangle}. Using
the Hilbert space defined by the Heitler-London approximation to the
pocket-states, leads directly to the Heisenberg model for this
system, with a spin-1/2 ground state, as shown in section~\ref{hubm}.
Ring terms corresponding to cyclic rotations of all
three electrons are not included in this model. However, it is
clear from the pocket-state description that such processes
exist. The relevant tunneling integrals between the $\:3!=6\:$
pocket-states in this case can be represented by permutations
of the types
\bea
(123)&\longrightarrow& (213)\quad :\quad J/2\qquad
\mbox{Pair Exchange}\\[3ex]
(123)&\longrightarrow& (312)\quad :\quad K/2\qquad
\mbox{Ring Exchange}\quad.
\eea
Their magnitudes may be estimated semiclassically within the WKB
approximation. The most important contributions to $\:J\:$ and
$\:K\:$ vary exponentially with the particle distance $\:r_{\rm s}\:$
(which equals the length of one side of the triangle in the case
considered), i.e.
\[
J/2\sim {\rm e}^{-\mbox{\footnotesize\S}_{(123)\to (213)}}\quad,\qquad
K/2\sim {\rm e}^{-\mbox{\footnotesize\S}_{(123)\to (312)}}
\]
where
\[
\S=\int_0^1{\rm d}q\;\sqrt{2mnW(\vec{\bm{x}}(q))}
\]
is the action associated with the transport of $\:n\:$ particles
($\:n=2\:$ or $\:n=3\:$ for $\:J\:$ or $\:K\:$, respectively) of
mass $\:m\:$ from the initial classical ground state $\:(123)\:$
into the final state. The transition takes place along the
trajectory $\:\vec{\bm{x}}(q)\:$ which in principle obeys an
equation of motion which minimises the classical action.
Here, it is parametrized by $\:0\le q\le 1\:$ in
configuration space (of dimensionality $\:dN=6\:$). For an
explicit estimate let us specify the interaction between the
particles and assume $\:w(|\bm{x}_i-\bm{x}_j|)=\d\frac{e^2/\kappa}
{|\bm{x}_i-\bm{x}_j|}\:$. Then $\:W(\vec{\bm{x}})\:$ is the Coulomb
energy at $\:\vec{\bm{x}}\equiv\{\bm{x}_1,\bm{x}_2,\bm{x}_3\}\:$ where
the equilibrium value, $\:\d 3\frac{e^2/\kappa}{r_{\rm s}}\:$, has been
subtracted so that $\:W(\vec{\bm{x}}(0))=W(\vec{\bm{x}}(1))=0\:$.
The results
\bea
\S_{(123)\to (213)}&=&\d 4\sqrt{m\frac{e^2}{\kappa}r_{\rm s}}
\int_0^{1/2}{\rm d}q\;\left[
\frac{1}{\sqrt{1-q+q^2}}+\frac{1}{\sqrt{1-\frac{29}{8}q+4q^2}}+
\frac{1}{\sqrt{1-\frac{5}{2}q+\frac{7}{4}q^2}}-3
\right]^{1/2}\\[3ex]
&=&2.139\sqrt{r_{\rm s}me^2/\kappa}\\[3ex]
\S_{(123)\to (312)}&=&6 \d\sqrt{m\frac{e^2}{\kappa}r_{\rm s}}
\int_0^{1/2}{\rm d}q\;\left[
\frac{1}{\sqrt{1-3q+3q^2}}-1\right]^{1/2}\\[3ex]
&=&2.852\sqrt{r_{\rm s}me^2/\kappa}
\eea
are obtained assuming straight lines $\:\vec{\bm{x}}(q)\:$ for
simplicity. This yields upper bounds to the true values of \S\
which may be reasonable particularly for $\:\S_{(123)\to
(312)}\:$. The estimate for $\:\S_{(123)\to (213)}\:$ is surely
worse because the classical path $\:\vec{\bm{x}}(q)\:$ is more difficult
to estimate for the pair exchange. The chosen paths for both cases
are indicated in figure~\ref{triangle}. Thus the pair exchange
process (corresponding to the superexchange within the lattice
description below) is slightly dominant compared to the ring
process in the equilateral triangle. This supports a ground state
of low total spin $\:S=1/2\:$. However, the triangle can easily
be distorted so that $\:\S_{(123)\to (312)}\:$ is much reduced
while $\:\S_{(123)\to (213)}\:$ remains almost unaffected, resulting
in a crossover to a spin polarised ground state \cite{stevens,trieste}
(see also below).

The existence of ring exchange processes highlights a fundamental
difference between a true lattice model and a continuum model. In
order to accommodate such processes we need a better lattice
approximation to the continuum than that suggested by the Wigner
lattice. This may be achieved by introducing intermediate
lattice points and corresponding localised one-electron orbitals.
Orthogonalising these one-electron orbitals then enables an
orthonormal many-electron basis set to be constructed, as before.
This is shown in figure~\ref{addpoints} for the triangular dot in which
an extra lattice point is inserted between each pair of vertices.

The low-energy eigenstates may now be expanded in this basis set.
As before, the probability of occupation of the three vertices
is highest, leading to a $2^3=8$-fold degenerate ground-manifold
when the hopping terms are switched off. Higher-lying base
states, corresponding to double-occupation of a lattice site and
to occupation of intermediate lattice sites may then be
eliminated by degenerate perturbation theory, resulting once more
in an effective spin-Hamiltonian. Intermediate states which
involve double-occupation of a lattice site give rise to a
Heisenberg superexchange term in the effective Hamiltonian, with
$J=8t^4/V^2U$, where $t, V$ and $U$ have the same meanings as
before but now refer to nearest-neighbor and on-site
interactions for the new lattice. Longer-range interactions are
ignored. (Note, however, that this is not necessary, it being straightforward
to include longer-range interactions leading to a renormalisation
of $J$ which will leave the form of the (Heisenberg) pair
exchange unchanged.) There are two processes involved in this superexchange
making equal contributions to $J$. These are shown in figure~\ref{processes}.
These processes are analogous to superexchange in atomic systems
in which the intermediate sites would correspond to ligand ions
surrounding transition-metal ions. Recent examples of this kind
of system are the copper-oxide planes of the high-temperature
superconductors, for which there is a similar fourth-order
expression for the superexchange \cite{eskes}. It is also clear from
this new lattice description that other processes, which do not
involve double occupation of a site (and therefore do not involve
the exclusion principle), are potentially important. The lowest-order
process of this type is also a pair exchange and occurs in
fifth-order. An example is shown in figure~\ref{processes}~(b),
which makes a contribution to $J$ of order $t^5/V^4$. Despite being
higher-order than the other superexchange terms, this process is, in
fact, the largest since it does not involve double-occupation of
a site (i.e.\ there is no $U$-factor in the energy denominator).

With intermediate lattice points it is easy to see that there are
ring processes giving rise to simultaneous rotation of all three
electrons. One such process, in which the three electrons are
rotated cyclically by $2\pi/3$, is shown in figure~\ref{triproc}.
The process resembles the classical path which would contribute
to the WKB tunneling rate, cf.\ figure~\ref{triangle}. The former
is a sixth-order process giving rise to a cyclic rotation operator
in the effective Hamiltonian with amplitude $t^6/12V^5$. Summing over
all such processes results directly in an effective `ring' operator,
$-K(R_{2\pi/3}+R_{-2\pi/3})$, where $K$ is a constant energy and
$R_{2\pi/3}$ ($R_{-2\pi/3}$) is an operator which performs a cyclic
(anticyclic) rotation of all three electrons (c.f.\ (\ref{rotop})).
We note that these rotation operators may be written as spin-operators
using the identities,
\be\label{cycperm}
R_{2\pi/3}\equiv P_{12}P_{23} {\rm~and~}
R_{-2\pi/3}\equiv P_{23}P_{12}
\ee
where $P_{ij}\equiv 2\bm{s}_1\cdot\bm{s}_2+1/2$ is the Dirac
exchange operator. Hence, the effective spin-Hamiltonian for 3
electrons in an equilateral triangular potential well becomes,
\be\label{tri}
H_{\rm eff}=\frac{J}{2}\left[ (P_{12}-1)+(P_{23}-1)+(P_{31}-1)\right]
-K(P_{12}P_{23}+P_{23}P_{12})\quad.
\ee

We see immediately from this form that the spin polarised state
has eigenenergy $-2K$. It is straightforward to complete the
diagonalisation of $H_{\rm eff}$ which results in a further pair
of degenerate doublets at energy $K - 3J/2$. Thus we see that the
ring and exchange terms oppose each other with the former
favouring a high--spin ground state and the latter a low--spin
ground state. The crossover occurs at $J=2K$, which is not quite reached
for the equilateral triangle, as discussed above.

This will not always be so for other geometries as follows
immediately from the fact that we can change the shape of
the dot in such a way that the ring processes are increased in
amplitude relative to the exchange processes. An `extreme'
example of this is the circular dot for which there is no barrier
to the simultaneous rotation of the three electrons and the ground
state should be spin polarised. The transition from high-spin to
low-spin ground states in rings with increasing impurity barrier has been
discussed recently \cite{trieste} in the context of persistent currents.

With the inclusion of higher--order exchange processes involving
also more than two particles, this completes the mapping of formerly
continuous electron problems (\ref{genmodel}) onto lattice models
(\ref{tJV}) for their low energy and spin properties.
This incorporates all permutational processes which according
to the pocket-state description may be relevant. The
above derivation of an effective-spin model for a triangular dot,
including ring terms, is easily generalised to arbitrary shapes
by inspecting all possible classes of ring and exchange terms on
a suitably dense lattice into (\ref{tJV}). The relative
amplitudes of the various processes can be estimated
semiclassically.

\section{Comparison with exact results and other
approximations}\label{compare}
To corroborate the lattice description for the low energy and
spin properties of quantum dots we compare the results from
(\ref{tJV}) with results obtained by numerically exact
diagonalisations \cite{haus93} and by the pocket-state method
\cite{pocket}.

The simplest model for an ``artificial atom'' is a hard wall box
in one dimension. Rigorous properties for the sequence of spin
states
\be\label{LM}
E(S)<E(S')
\ee
for $\:S<S'\:$ are known by a theorem of Lieb and Mattis
\cite{liebmattis}. $\:E(S)\:$ denotes the lowest energy of $\:N\:$
almost arbitrarily interacting electrons to given spin
$\:S=\Kg{\d 0\atop\d 1/2},\ldots,N/2\:$. In particular, the
ground-state spin must be minimal. This rule (\ref{LM}) is in obvious
agreement with the description in terms of an antiferromagnetic
Heisenberg (spin--$1/2\:$) chain (\ref{hj}), valid for this case
according to section~\ref{hubm}, where the exchange of adjacent
electrons is the relevant permutational process and
there is only one classical minimum energy configuration,
($\:\nu=1\:$). In
table~\ref{upn4} the energy values of $\:N=2,3,4\:$
electrons are given in units of $\:J\:$ as obtained
within the pocket-state approximation (PSA), assuming
$\:J_{12}=J_{23}=J_{34}\:$. These spectra agree exactly with the
eigenvalues of the Heisenberg chain of $\:N=2,3,4\:$ spins apart
from unimportant additive constants. Similarly, figure~\ref{1dlevels} shows
the excitation energies for $\:N=3,\ldots,6\:$ and includes also
results obtained by numerical diagonalisation of the Hamiltonian
matrices for $\:N=3,4\:$ (dashed). For the cases $\:N=5\:$
and $\:N=6\:$ we also checked, by numerical diagonalisation, that the
Heisenberg chain shows exactly the same spectrum.

As a further example we investigated $\:N=2,\ldots,5\:$ electrons
in a two-dimensional square with hard walls. For these cases the
interaction must be long range $\:w(|\bm{x}|)\sim |x|^{-\gamma}\:$
with $\:\gamma<2\:$ in order to make the PSA
applicable. The spectra obtained within PSA are shown in
figure~\ref{2dlevels}. For $\:N=4\:$ the nearest neighbor pair
exchange process has been compared with the competing process of
all 4 electrons rotating cyclically by $\:\pi/4\:$ \cite{pocket},
in a similar way as it has been done for $N=3$ electrons in a
triangle in the previous section,
with the result of the former being slightly dominant. In all
cases the equilibrium positions of the electrons have been taken
as sites for single electron states for the lattice
description, as discussed in section~\ref{hubm}, and the importance
of transition processes has been estimated semiclassically.
For $\:N=5\:$ one electron is, by electrostatics, located in
the centre of the square being coupled antiferromagnetically,
according to (\ref{hj}), to the 4 electrons placed in the
corners. The Heisenberg model is appropriate since there is only one
classical equilibrium configuration ($\:\nu=1\:$). Neglecting next
nearest neighbor exchange results in a $\:S=3/2\:$ ground state
spin, using the PSA and permutation group theory. This result is
immediately understandable from the Heisenberg spin-Hamiltonian, which
clearly has a quartet ground state. Furthermore, we have checked that
for all spectra, the corresponding
Heisenberg ($\:N=4,5\:$) or $t-J-V$ ($\:N=2,3\:$) Hamiltonians exactly
reproduce the correct ordering of the spin multiplets and the correct
energy splittings provided that the parameters $t,J$ and $V$ are
chosen to be consistent with the corresponding tunnelling amplitudes
in the PSA.

As a third example we look at the spectrum of three electrons in a
one-dimensional ring. In section~\ref{ring} we explained why the
ground state is high spin ($\:S=3/2\:$), which is also found in the
pocket-state description. Two degenerate low spin ($\:S=1/2\:$)
excited states are higher in energy by the rotational constant
$\:h^2/2mNL^2\:$, where $\:L\:$ is the circumference of the ring.
They are additionally shifted by the magnitude of the pair exchange
process $\:J\:$ towards lower energy, in agreement with
(\ref{tri}).

\section{Summary and outlook}\label{outlook}
Based on the pocket-state description of the low energy
excitations in finite and a priori {\em continuous} systems of
interacting electrons at low densities we have derived a lattice
formulation for this problem. Non--orthogonal single electron
states centred near the positions where the electrons would be in
the classical minimum energy configurations are orthogonalised,
leading to contributions from double occupancies $\:\sim U_i
n_{i\uparrow}n_{i\downarrow}\:$ in an effective Hubbard model
description. Subsequent elimination of high energy states ($\:\sim
U_i\:$ or $\:\sim V_{ij}\:$, from nearest neighbor occupations)
yields effective antiferromagnetic Heisenberg (eq.~(\ref{hj}) for
$\:\nu=1\:$) or charge-spin Hamiltonians (eq.~(\ref{tJV}) for
$\:\nu>1\:$) depending on the number $\:\nu\:$ of classical
minimum energy configurations. An extended version of the Hubbard
model must be considered if highly symmetric geometries or ring
processes are important. A crucial enhancement has been to
enable ``ring exchange'' processes to be included, corresponding
to the cyclic permutation of $\:N>2\:$ electrons. Intermediate
lattice points were introduced and the additional states corresponding
to their single occupation were also eliminated by higher order
perturbation theory, even though they are much lower in
energy than states corresponding to double occupation. This
yields generalized effective spin-1/2 models containing up to
$\:(N-1)$--fold products of pairs of spin operators (see
Eqn.~\ref{cycperm}). The magnitudes of the parameters
for the superexchanges $\:J\:$ and the ring exchanges $\:K\:$
can be estimated semiclassically.

Apart from providing a quantitative description of the low-energy
spectra in quantum dots the mapping onto lattice models is, in itself,
very illuminating. For example, in cases that can be mapped
onto the antiferromagnetic Heisenberg model for the low energy states,
the spectrum consists of a level multiplet of `band width'
$\:NJ\:$ where $\:J\:$, the nearest neighbor
superexchange integral, depends exponentially
on the mean electron distance $\:r_{\rm s}\:$. The
highest energy state within this multiplet is the spin
polarized state, in agreement with what has been derived within
the pocket-state approximation using less obvious arguments. All
spectra obtained by the lattice descriptions coincide exactly
with those found in the pocket-state \cite{pocket} approximation and are
in very convincing agreement with results gained by numerical
diagonalisations \cite{haus93,bryant}. The computational effort required
for the former is, however, considerably less compared to both of the
other methods. This will enable much larger systems with $N>20$ to be
investigated.

In certain geometries of two dimensional quantum dots, $\:\nu>1\:$
classical configurations of lowest electrostatic energy may
exist leading to an effective spin-charge model instead of
a spin only effective model.
Whilst such situations are straightforward to deal
with by the lattice method, we should point out that in real
experimental situations the high symmetry will be lowered by a
polarizable environment for the
following reason. The energetically most favourable locations
for the quantum dot electrons depend on the distribution of
surrounding (non--conducting) charges, which themselves are
influenced by the distribution of the (Wigner localised but
under transport conditions conducting) dot charges.
Selfconsistency allows the dot electrons to lower the
total energy by adjusting their environment. For example, two
electrons in a square will easily break the two-fold symmetry of the
classical ground-states by polarizing their surroundings,
leaving a diamond shaped configuration for the potential. The
interplay between the surroundings and the granular electron
density of the dot finally tends to lower the number of classical
low energy configurations. Only configurations with energies
smaller than the band width $\:\sim NJ\:$ eventually need to
be considered explicitly. This plays a particular role e.g.\
in double-dot systems \cite{doubled}. Additionally,
one should keep in mind that the shape of the quantum dot
also depends on voltages applied to side gates \cite{leo},
so that the fine structure spectra may change with gate voltage.

The lattice description simplifies considerably the determination
of excitations in quantum dots. These excited levels may be deduced
implicitly from non-linear transport experiments \cite{korotkov}.
Furthermore, transport is qualitatively influenced by the total spins
of the many-electron states by two types of {\em spin blockades}
\cite{weinmann94,weinmann95}.

The occurrence of a non-minimal total spin for the ground state
in a finite electron system is
a direct consequence of the geometry of the boundary. For systems
which can be represented by an antiferromagnetic Heisenberg model,
this relationship is obvious (c.f. discussion of the square dot with $N=5$
in section~\ref{compare}). Apart from the electron number the spin values
of ground and of excited states should depend very sensitively on the
shape of the quantum dot. In future research this will be investigated
systematically to allow specific suggestions for experimental
design.

Another highly relevant topic for real experimental situations is
to apply the mapping derived in the present work to study the
interplay between mutual interactions of the quantum dot electrons
and impurities. No reliable picture exists presently, which demonstrates
our lack of understanding for the persistent currents observed in
mesoscopic rings. Theory is still not capable of explaining even
the order of magnitude of the high values found experimentally.
It is known that one dimensional models of rings are insufficient
to describe this problem \cite{berkovits95}. Furthermore, evidence exists
that the electron spin is important \cite{giamarchi,trieste}. Our mapping
onto lattice models demonstrates that calculations based on the Hubbard
model \cite{fowler,ramin95,romer} are not significantly influenced by the
lattice but indeed
reproduce the low energy properties of the original continuous problems
in presence of strong interactions correctly if the parameters,
particularly the filling of the former, are interpreted accordingly.
Advantage can be taken of the comparably large electron numbers that
can be dealt with by our method, which would be otherwise intractable.
 This will allow the investigation of
multiple-channels in two-dimensional rings of finite widths in
the presence of impurities and flux.

\vspace{1cm}\noindent {\bf Acknowledgements}\\

We acknowledge stimulating discussions with Bernhard Kramer,
Colin Lambert, Walter Stephan and colleagues in our EU-sponsored HCM network
on the quantum dynamics of phase coherent structures.
(HCM No. CHRX-CT93-0136)
\newcommand{\pap}[5]{#1, \ #2 {\bf #3}, #4 (19#5)}
\newcommand{\bk}[4]{#1 \ {\it ``#2''},\\ #3 (19#4)}

\noindent\begin{table}[h]
\begin{center}
\begin{tabular}{c@{\hspace{9mm}}c@{\hspace{9mm}}c}
$N$&$S$&$E_m^{\mbox{\tiny\rm $(N)$}}-
E_{\mbox{\tiny\rm Ground state}}^{\mbox{\tiny\rm $(N)$}}$\\ \hline
$2$&$0$&$0$\\
$2$&$1$&$J^{\mbox{\tiny\rm $(2)$}}$\\
$3$&$1/2$&$0$\\
$3$&$1/2$&$J^{\mbox{\tiny\rm $(3)$}}$\\
$3$&$3/2$&$(3/2)J^{\mbox{\tiny\rm $(3)$}}$\\
$4$&$0$&$0$\\
$4$&$1$&$(1+\sqrt{3}-\sqrt{2})J^{\mbox{\tiny\rm $(4)$}}/2$\\
$4$&$1$&$(1+\sqrt{3})J^{\mbox{\tiny\rm $(4)$}}/2$\\
$4$&$0$&$(2\sqrt{3})J^{\mbox{\tiny\rm $(4)$}}/2$\\
$4$&$1$&$(1+\sqrt{3}+\sqrt{2})J^{\mbox{\tiny\rm $(4)$}}/2$\\
$4$&$2$&$(3+\sqrt{3})J^{\mbox{\tiny\rm $(4)$}}/2$
\end{tabular}
\end{center}
\parbox[b]{14cm}
{\caption[upn4]{\label{upn4}
Analytical values for the fine structure spectrum
$\:E_m^{\mbox{\tiny\rm $(N)$}}\:$ of $\:N\:$ interacting
electrons in a one-dimensional hard wall box
within PSA for $\:N\le 4\:$. $\:S\:$ is the total spin of $\:N\:$
Fermions with $\:s=1/2\:$. The excitation energies
$\:E_m^{\mbox{\tiny\rm $(N)$}}-E_{\mbox{\tiny\rm
Ground state}}^{\mbox{\tiny\rm $(N)$}}\:$, are
given in units of $\:J^{\mbox{\tiny\rm $(N)$}}\:$.
}}
\end{table}

\noindent
\begin{figure}
\caption{Classical low-energy configurations for two electrons in a 2D square
quantum well. The first two configurations are lowest in energy
whilst the remaining four are $V$ higher in energy due to Coulomb
repulsion. All other configurations have higher electrostatic energy.}
\label{squararr}
\end{figure}

\noindent\begin{figure}
\caption{(a) Classical ground-state configuration for three electrons in a
triangular quantum dot. (b) Classical path for the exchange of two
electrons. (c) Classical path for the cyclic permutation of all three
electrons. Particles are shown in intermediate positions along
their trajectories.}
\label{triangle}
\end{figure}

\noindent\begin{figure}
\caption{Lattice points for the triangular quantum dot. In the ground
manifold all three electrons will be close to the positions indicated
by the circles. Excited (intermediate) states correspond to one or
more electrons in localised orbitals centered on the crosses.}
\label{addpoints}
\end{figure}

\noindent\begin{figure}
\caption{Superexchange processes. (a) Fourth-order ($\sim t^4/V^2U$).
(b) Fifth-order ($\sim t^5/V^4$). }
\label{processes}
\end{figure}

\noindent\begin{figure}
\caption{A six-step ring process for three electrons in a triangular
quantum dot. The electrons would move continuously and simultaneously
in the direction of the arrows in the WKB approximation. }
\label{triproc}
\end{figure}

\noindent\begin{figure}
\caption{Fine structure multiplets of a quasi one-dimensional
quantum dot, for $\:N=3,\ldots,6\:$ as obtained
directly from the pocket-state approximation and in exact agreement with
the eigensolutions of the effective spin Hamiltonian. The dashed lines were
obtained by direct numerical diagonalisation of the interacting electron
problem. The $\:N$--dependence of $\:J^{(N)}\:$
is not considered and $\:J\:$
has been adjusted to normalize the `bandwidth' of the
multiplets.}
\label{1dlevels}
\end{figure}

\noindent\begin{figure}
\caption{Fine structure spectra of (a) $\:N=3\:$, (b) $\:N=4\:$ and
(c) $\:N=5\:$ electrons in a 2D square as obtained within the
pocket-state approximation. In (a) the dominant tunneling process
($t$) corresponds to nearest-neighbor hopping of an electron whereas
in (b) and (c) the dominant tunneling process is nearest-neighbor exchange
($J$). All spectra are identical with those obtained from the corresponding
charge-spin Hamiltonian.}
\label{2dlevels}
\end{figure}


\begin{thebibliography}{10}
\bibitem{footnote}It is possible to
include the $V$-term in the unperturbed Hamiltonian. However,
this introduces complications in the perturbation expansions
(since the intermediate state energies depend on their
environment) which are best avoided. Since we are interested in
the regime for which $U\gg V$, including $V$ in the perturbation
introduces only a small error.
\end{thebibliography}
\end{document}